# A possible Solution of the P versus NP Problem based on specific property of clique function

Boyu Sima
Nanjing University of Science and Technology
No. 200 Xiao Lingwei Street, Nanjing, China
Email: smby@njust.edu.cn

**Abstract**

Circuit lower bounds are important since it is believed that a super-polynomial circuit lower bound for a problem in NP implies that P≠NP. Razborov has proved superpolynomial lower bounds for monotone circuits by using "method of approximation". However, until now, no one could prove a non-linear lower bound for the non-monotone complexity of any Boolean function in NP. We show that by replacement of each "Not" gates into constant "1" equivalently in standard circuit for clique problem, it can be proved that non-monotone network has the same or higher lower bound compared to the monotone one for computing the clique function. This indicates that the non-monotone network complexity of the clique function is super-polynomial which implies that P≠NP.

**Introduction**

An attempt to solve P versus NP Problem is to demonstrate whether a super-polynomial lower bound on the size of Boolean circuits solving NP-complete problem, like 3-SAT or Clique problem exists. In 1985, Razborov [1] and Andreev [2] successively proved an $n^{\Omega(\log n)}$ lower bound on the monotone-size of the clique function by using "method of approximation". This was the first super-polynomial bound on the monotone-size of any explicit function and was improved to $\omega(n^k/\log^k n)$ by Alon and Boppana [3] later. Some other works also proved super-polynomial lower bound on the monotone-size of clique-like functions with similar approach but more beautiful presentations [4,5]. But no one could prove a non-linear lower bound for the non-monotone complexity of any Boolean function in NP in the past few decades.

In the paper "A solution of the P versus NP problem based on specific property of clique function" [6], we demonstrated that non-monotone complexity of the clique function is equal to or even larger than the monotone complexity by considering specific property of the Clique function. While Brendon Pon [7] claim that our argument is incorrect due to the reason that connection between a Boolean variable and its negation is overlooked during the process of proof. He also presented a simple example for demonstration.

Thanks for his concern. The problem he found does exist. However, it can be easily fixed by adding some clarifications. And the example he presented actually does not conflict with our basic argument which is "Every non-monotone circuit computing clique function can be equivalently transformed to a monotone circuit without increasing complexity". Therefore, the main strategy used in our original paper is still correct. In this paper, we will renew the poof procedure with more clear descriptions to fix the loophole found by Brendon Pon and show the validity of our original method.

**Preliminaries**

A Boolean circuit is a directed acyclic graph with gate nodes (or, simply gates) and input nodes. Operation AND or OR is associated with each gate whose indegree is 2 which is represented by "$\wedge$" and "$\vee$" in this paper for short, respectively. Not gate whose in-degree is 1 which is represented by "$\neg$" for short, like "$\neg A$" represents for NOT($A$). A Boolean variable or a constant, namely, 0 or 1, is associated with each input node whose in-degree is 0. In particular, a circuit with no NOT gates is called monotone. A Boolean function of n variables is called monotone if $f(w_1) \leq f(w_2)$ holds for any $w_1, w_2 \in \{0,1\}^n$ such that $w_1 \leq w_2$. Let $M_n$ denote the set of all monotone functions of n variables. The size of a circuit C, denoted size(C), is the number of gates in the circuit C. The circuit complexity of a function $f$, denoted by size($f$), is the size of the smallest circuit computing $f$.

For 1<=s<=m, let CLIQUE(m,s)(x) be the Boolean function of $n := \binom{m}{2}$ variables representing the edges of an undirected graph G = (V, E) on m nodes ($x = (x_1, x_2...x_n) \in \{0,1\}^n$ represent for input variables. When the value of a variable is 1 means the corresponding edge is connected and vice versa). CLIQUE(m,s)(x) = 1 iff the corresponding graph G contains a clique of size s. Let $\beta_c(x)$ denote the circuit which compute the Boolean function CLIQUE(m,s)(x).

For any circuit network $\beta$, we can convert $\beta$ to an equivalent network $\beta_{st}$ where all negations occur only at the input nodes and the size of $\beta$ is at most doubled. The equivalent network $\beta_{st}$ is a so-called standard network where only input variables are negated. We consider a negated variable $\neg x_1$ as an input node g with op(g) = $\neg x_1$. The standard circuit complexity $C_{st}(f)$ of a function $f \in$ Mn is the size of a smallest standard network which computes $f$. Note that the standard and the non-monotone complexity of a function $f$ differs at most by the factor two. Hence, for proving a super-linear lower bound for the non-monotone complexity of a Boolean function, we can restrict us to the consideration of standard networks.

Now we can suppose that the $\beta_{cst}(x)$ which compute the Boolean function CLIQUE(*m*,*s*)(*x*) can be written as:

$$res_{\beta_{cst}}(x) = CLIQUE(m,s)(x) \tag{1}$$

Where $x = (x_1, x_2 ... x_n) \in \{0,1\}^n$ are input variables. $res_{\beta_{cst}}(x)$ represents the output of the whole circuit. For the reason that $\beta_{cst}(x)$ is a standard circuit network, it always can be written as a DNF (disjunctive normal form) formula like shown below:

$$res_{\beta_{cst}}(x) = \bigvee_{i=1}^{t} m_i \tag{2}$$

Where each $m_i$ is a monomial (conjunction of some literals) which can be represented as:

$$m_i = x_1^{i(1)} \wedge \cdots \wedge x_n^{i(n)} \wedge (\neg x)_1^{i(n+1)} \wedge \cdots \wedge (\neg x)_n^{i(2n)} \tag{3}$$

Where $i(j) \in \{0,1\}$, $1 \le j \le 2n$, when $i(j) = 1$ it means monomial $m_i$ contains corresponding variable and when $i(j) = 0$ it means monomial $m_i$ does not contain corresponding variable.

## Proofs of the equivalence of the monotone and non-monotone complexity for the clique function

It has been proved that the lower bounds for the monotone network complexity of the clique function is exponential. Now we will demonstrate that any non-monotone network with "NOT" gate for the clique function can be transformed to an equivalent monotone circuit without increment of the circuit size.

According to Distribution Law, we can get the equation below by exacting the negated variable $\neg x_1$ from $\beta_{cst}(x)$:

$$
\begin{aligned}
res_{\beta_{cst}}(x) &= \left( \bigvee_{a=r1(1)}^{r1(w)} m_a \right) \vee \left( \bigvee_{b=r2(1)}^{r2(t-w)} m_b \right) \quad (4)\\
&= \left[ (\neg x_1) \wedge p \right] \vee \left( \bigvee_{k2=r(1)}^{r2(t-s)} m_{k2} \right)
\end{aligned}
$$

Where $m_a$ is monomial with negative variable $\neg x_1$, $m_b$ is monomial without containing negative variable $\neg x_1$. It should be noted that all $m_a$ do not contain variable $x_1$. Because $\neg x_1$ and $x_1$ can not exist in the same monomial, other vice the monomial will be a constant value of "0" which dose not have any meanings. $p$ is the rest part of $\bigvee_{a=r1(1)}^{r1(w)} m_a$ with negative variable $\neg x_1$ exacted out. According to the analysis above, $p$ is a disjunctive normal form (DNF) containing neither $\neg x_1$ nor $x_1$. Therefore, $p$ is independent with both $\neg x_1$ and $x_1$. This is very important which is not involved in the previous paper. The loophole found by Brendon Pon is just due to the lack of this illustration.

Then the theorem below will demonstrate that the negated variable $\neg x_1$ can be replaced by constant 1 without influence the value of $res_{\beta_{cst}}(x)$ for Clique function with all different inputs. This means the transform is equivalent with negated variable $\neg x_1$ replaced by constant 1.

**Theorem 1** Let $\beta_{cst}(x)$ be a standard network which computes CLIQUE(m,s)(x) Boolean function. Then the following hold:

By replacement of one of the negated variables $\overline{x_i}(i=1...n)$ in $\beta_{cst}(x)$ into constant 1, the new network $\beta_{cst}'(x)$ still computes the same clique function CLIQUE(m,s) (x) correctly.

**Proof of Theorem 1:**

Let's focus on the first term in (4) which is $\left[(\neg x_1) \wedge p\right]$. This is the only term containing $\neg x_1$. By considering the characteristics of the clique function, we can analysis the influence of the replacement of $\neg x_1$ to constant 1. As can be seen directly, $\left[(\neg x_1) \wedge p\right]$ consists of two part which is $\neg x_1$ and $p$, respectively. According to the value of $p$, we distinguish two cases.

Case 1: The value of $p$ is 0.

It is obvious that when the value of $p$ is 0, the replacement of $\neg x_1$ to constant 1 will not have any influence to the value of $\left[(\neg x_1) \wedge p\right]$( first term in (4)). Because no matter what the value of $\neg x_1$ is, the value of $\left[(\neg x_1) \wedge p\right]$ will always be 0.

Case 2: The value of $p$ is 1.

This is the key part of the proposed method. We will demonstrate that when the value of $p$ is 1, it will lead to $res_{\beta_{cst}}(x) = 1$ (that means the corresponding graph G under this condition contains a clique of size s) no matter what the value of

$\neg x_1$ is. Suppose $p=1$ and $\neg x_1 = 1$, than value of $[(\neg x_1) \wedge p]$ being 1 which leads to $res_{\beta_{cst}}(x) = 1$ according to (4). However, $\neg x_1$ equals to 1, that is $x_1$ equals to 0 means the corresponding edge is disconnected according the definition of the clique function. This implies that the edge that $x_1$ stands for has no contribution to the size of clique. Thus it can be concluded that the "1" value of $\beta_{cst}(x)$ comes totally from other variables. The connections of other edges contribute to the existence of a clique of size s and obviously this clique do not contain the edge of $x_1$. Thus, the value of $\beta_{cst}(x)$ being "1" just because $p=1$ which has no relationship with the value of $\neg x_1$. Thus $\neg x_1$ can also be replaced by constant 1 without influence on the output of circuit when the value of $p=1$ is 1.

Considering both situations, we have proved the Theorem 1.

Then, we extend the replacement to all of negated variables one by one. The following Theorem 2 can be naturally proved.

**Theorem 2** Let $\beta_{cst}(x)$ be a standard network which computes CLIQUE(m,s)(x) function. Then By replacement of all of the negated variables in $\beta_{cst}(x)$ into contant 1, the new network $\beta_{cst}"(x)$ still computes CLIQUE(m,s)(x) function correctly.

**Proof of Theorem 2:**

According to Theorem 1, we can have:

$$\beta_{cst}' = \beta_{cst}(\neg x_1 \Rightarrow 1) = \beta_{cst} \tag{5}$$

Where $\beta_{cst}(\neg x_1 \Rightarrow 1)$ represents performing the equivalent transform of $\neg x_1$. $\beta_{cst}'$ is the new circuit after equivalent transform whose input variables contain $(x_1, x_2, ... \neg x_2, ..., \neg x_n)$ except $\neg x_1$.

We can repeat this equivalent transform to all the negative variables one by one. At last, we can get $\beta_{cst}''(x)$ with all negative variables transformed to constant "1" equivalently. Because there is not any negative variable in its input nodes, $\beta_{cst}''(x)$ is a monotone circuit network that computes the same clique function as $\beta_{cst}(x)$ dose.

Theorem 2 indicates that any standard network which computes CLIQUE(m,s)(x) Boolean function can be transformed to an equivalent monotone circuit by replacement of all the negated variables to constant 1. It is obvious that this process will not increase the complexity of the circuit. This means that standard network do not have smaller circuit size than the monotone one for Clique function. For the reason that the circuit size of monotone network of Clique function has proven to be exponential, we can conclude that non-monotone network complexity of the clique function is also super-polynomial which implies that P≠NP.

**Response to Critique from Brendon Pon**

In [7] Brendon Pon argued that our method is not satisfactory due to failing to consider the connection between a Boolean variable and its negation. The main reason is that he think we fail to consider the connection between *Term1part1* and *Term1part2* in our original paper. He also provide an example where *Term1part1* is set to be $\neg x_1$ and *Term1part2* to be $x_1$. In that case, an obvious mistaken can be found if using our method. The transform of $\neg x_1$ is no longer equivalent. But actually this loophole can be fixed easily by adding some more detailed illustrations. In this paper, we redefine the *Term1part1* and *Term1part2* by using DNF formula as shown in Eq. (4). *Term1part1* is still $\neg x_1$ and *Term1part2* become $p$ which is a disjunctive normal form (DNF) containing neither $\neg x_1$ nor $x_1$. In this way, $\neg x_1$ and $p$ actually are independent with each other. So that the loophole has been fixed. One may think that this is because we assume that $p$ do not contain $x_1$. First, we believe that it is reasonable to make this assumption as we claimed above. If $p$ contains $x_1$ there is no real meanings. Second, we can also assume $p$ may contain $x_1$ just like in the example that Brendon Pon provide. In fact, this behavior does not conflict with our conclusion as well. It should be noted that our basic argument is that any non-monotone network with "NOT" gate for the clique function can be transformed to an equivalent monotone circuit without increment of the circuit size. Just take the example provided by

Brendon Pon into consideration, we can make a simplification at first. Because the *Term1part1* is set to $\neg x_1$ and *Term1part2* is $x_1$, the *Term1* will become constant "0" after simplification ($\neg x_1 \wedge x_1 = 0$). You can find that $\neg x_1$ has already been eliminated. We even do not need to perform constant value transformation. The simplified network do not have $\neg x_1$ anymore and is still equivalent to the original one. This is actually consistent with our basic claim.

**Conclusion**

To sum up, the basic idea we use in the previous paper is still effective. To show it more clearly, we rewrite the proof process. The main difference is that we transform standard network into DNF formula at first. So that we do not need to consider the connection between Boolean variable and its negation which was regarded as a loophole we didn't realize in Brendon Pon's critique. By making this improvement, the method we proposed is still reasonable.

**References**

[1] Alexander A. Razborov. Lower bounds on the monotone complexity of some boolean functions. Doklady Akademii Nauk SSSR, 281:798–801, 1985. English translation in Soviet Math. Doklady 31 (1985), 354–357.

[2] A. Andreev, On a method for obtaining lower bounds for the complexity of individual monotone functions, Dolk. Akad. Nauk. SSSR 282(5) (1985), pp. 1033-1037 (in Russian). English translation in: Soviet Math. Dokl. 31(3) (1985), 530-534.

[3] N. Alon and Ravi B. Boppana. The monotone circuit complexity of boolean functions. Combinatorica, 7(1):1–22(1987).

[4] C. Berg, S. Ulfberg, Symmetric approximation arguments for monotone lower bounds without sunflowers, Comput. Complex. 8 (1999), 1–20.

[5] K. Amano, A. Maruoka, The potential of the approximation method, SIAM J. Comput. 33 (2004), 433–447.

[6] Boyu Sima, A solution of the P versus NP problem based on specific property of clique function. Technical Report arXiv:1911.00722 [cs.CC], Computing Research Repository, arXiv.org/corr/ (2019).

[7] Brendon Pon, Critique of Boyu Sima's Proof that P $\neq$ NP (2020). arXiv:2005.03256 [cs.CC].